# Clean-Slate Design of Next-Generation Optical Access

**Kyeong Soo Kim[1], Karin Ennser[1], Yogesh K. Dwivedi[2]**

[1] *College of Engineering, Swansea University, Swansea SA2 8PP, Wales, United Kingdom*
[2] *School of Business & Economics, Swansea University, Swansea SA2 8PP, Wales, United Kingdom*
*e-mail: {k.s.kim, k.ennser, y.k.dwivedi}@swansea.ac.uk*

**ABSTRACT**
We report the current status of our research on the clean-slate design of next-generation optical access (NGOA). We have been studying candidate architectures with a major focus on their elasticity to user demands, energy efficiency, and support of better Quality of Experience (QoE). One of the major challenges in this study is to establish a comparative analysis framework where we can assess the performances of candidate architectures in an objective and quantifiable way. In this paper we describe our efforts to meet this challenge: (1) the development of a new comparison framework based on integrated QoE and statistical hypothesis testing and (2) the implementation of a virtual test bed capturing important aspects from physical layer to application layer to end-user behaviour governing traffic generation. The comparison framework and the virtual test bed will provide researchers a sound basis and useful tools for comparative analysis in the clean-slate design of NGOA.

## 1. INTRODUCTION

There is an ever growing need and demand for higher bandwidth and better Quality of Service (QoS) in the Information and Communications Technology (ICT). Especially the Internet, on which many services are based, is expected to become more pervasive, connecting not only people but also sundry devices anywhere anytime. With such growth and increase in Internet-reliant services, comes the rising demand for energy in the underlying network infrastructure. The energy efficiency, therefore, becomes an important issue in designing Next-Generation Optical Access (NGOA) together with bandwidth and QoS support [1].

Like the clean-slate design of new Internet architecture [2], we have been studying NGOA architectures from scratch and not limited by the backward compatibility with existing optical access architectures, i.e., Time Division Multiplexing-Passive Optical Networks (TDM-PONs). The focus of our study is on the following important aspects of network architectures:

− **Elasticity**: This means the ability to manage overall performances to a certain level by fast provisioning of network resources based on user demands. We can learn much from the Cloud Computing revolution in this regard, which has been enabled by the integration and sharing of large-scale computing resources in globally located data centers through virtualization technology [3].
− **Energy Efficiency**: As mentioned, we require NGOA to be energy efficient. Therefore, a network architecture should be designed so that it can reduce the number of active components and systems by consolidating and sharing them over many users.
− **Quality of Experience (QoE)**: To quantify the benefits of designed NGOA architectures, we focus on the performances perceived by end-users (e.g., the session delay in web browsing and the quality of streaming video) rather than traditional network-level performance measures (e.g., packet delay and packet loss rate).

In this paper, we report the current status of our research on the clean-slate design of NGOA to address the issues above. The rest of the paper is organized as follows: In Section 2, we discuss some of major requirements for future NGOA architectures. In Section 3, we describe our work on the development of a new comparison framework where we can assess the performances of candidate architectures in an objective and quantifiable way, and the implementation of a virtual test bed capturing the important aspects from physical layer to application layer to end-user behaviour governing traffic generation. Section 4 summarizes our discussions in this paper.

## 2. REQUIREMENTS FOR NGOA ARCHITECTURES

### 2.1 Integration with Metro Network

Figure 1 shows the evolution of future network architecture from separate access and metro networks into integrated one. There come two immediate benefits from this integration. First, we can greatly reduce the Operational Expenditure (OPEX) and Capital Expenditure (CAPEX) by consolidating multiple Optical Line Terminals (OLTs) and Central Offices (COs) in the field. Second, due to different usage patterns between residential and business users, we can assign network resources more efficiently based on their combined usage patterns which tend to smooth out [4].

Long-Reach PON (LR-PON) has been extensively studied as a cost-effective solution for the access/metro integration because it extends the coverage of PON from traditional 20 km to 100+ km. Most of LR-PONs demonstrated are based on Wavelength Division Multiplexing (WDM) overlay of multiple TDM-PONs [5]. Note that it is easier to provide backward compatibility with the existing TDM-PONs in this case because traditional Time-Division Multiple Access (TDMA) scheme is used within each wavelength. However, the allocation of wavelength is rather static, and we cannot exploit the benefit of combined, packet-level resource assignment in both time and wavelength domains. In this regard, hybrid TDM/WDM-PONs based on tunable transceivers, like SUCCESS-HPON [4], would be a better candidate for the clean-slate design of NGOA where the backward compatibility is not a requirement.

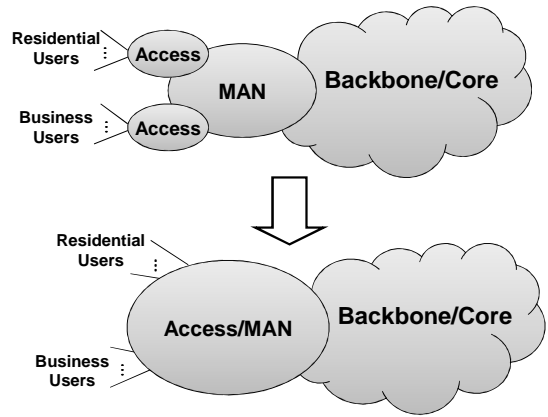

*Figure 1. Future integration of access and metro networks.*

### 2.2 Flexible Resource Sharing

Related with the access/metro integration discussed in Section 2.1, flexible resource sharing in multiple dimensions over many users is a key to both the elasticity and energy efficiency of network architecture. Figure 2 illustrates the resource sharing in point-to-point architecture (including static WDM-PON), TDM-PON, and hybrid TDM/WDM-PON. As indicated by a dotted ellipse, a transceiver at the Optical Line Terminal (OLT) is dedicated to one Optical Network Unit (ONU) in point-to-point architecture. In case of TDM-PON, as indicated by a dotted circle, it is shared by all the ONUs in a tree (i.e., PON), but resource sharing beyond a tree is still not possible, even in case when all the TDM-PONs are physically collocated through WDM overlay as in LR-PON. In hybrid TDM/WDM-PON, on the other hand, there exists no such a boundary in resource sharing; a pool of tunable transceivers at the OLT can be allocated to any ONUs in the network at a packet level. This not only maximizes resource sharing but also provides higher elasticity and energy efficiency [6].

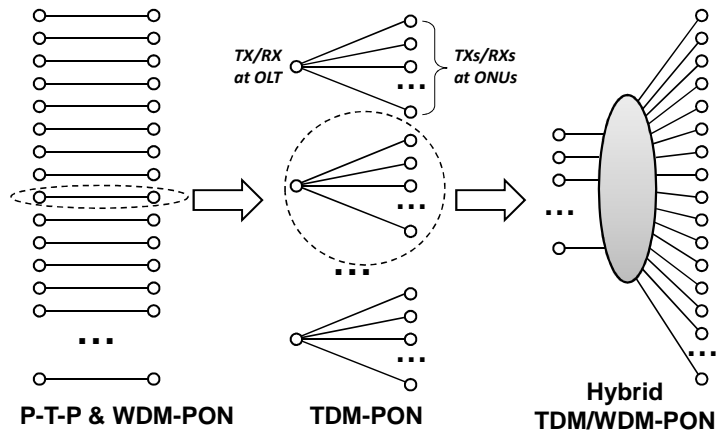

*Figure 2. Resource sharing in various network architectures.*

### 2.3 Higher Energy Efficiency

The consequence of burning fossil fuels, climate change and greenhouse gas emissions, are having an irreversible negative impact on the environment. Through international and national targets it is apparent a mutual understanding that there is a need for sustainable energy sources and energy consumption reduction. And hopefully meet the challenges such as the decreasing fossil fuel supply and minimizing $CO_2$ pollution and its environmental impacts. ICT plays dual parts in this matter. With the introduction of virtualization, information transmissions and teleconferencing, ICT prevents from transportation being utilized for documents and people and consequently reducing the Green House Gas (GHG) emissions. Essentially, ICT emits carbon footprint indirectly by consuming electricity part of which is produced by burning fossil fuels. According to Gartner's report [7] in 2007, the ICT industry accounts for two percent of all GHG emissions around the world. By 2020, the sector's GHG output may increase to as much as 1.4 billion tons of carbon dioxide equivalent.

According to a recent study [8], more than 50 percent of enterprises consider "greenness" when selecting a vendor and nearly 80 percent of executives say Green IT is becoming more important to their organizations. So Companies will soon have to deal with only Green conscious service providers therefore the need and urgency for the "greenness" in the access network. One of the major drivers of Green IT is the economical aspect of energy consumption saving. Cutting back power consumption has a direct impact on OPEX by reducing energy costs and also on CAPEX by minimizing the amount of equipment needed for cooling and backup power.

In a power consumption breakdown of the (1G) class optical access system [9], the power used by the core network is less than 20% of the total system whereas the remaining 80% is consumed by the access network. At the access network (PON), the power consumption is shared as 14% at layer 2 switches, 7% at OLT and about

60% at ONU. Emigrating to 10G class services will increase greatly the existing usage. Therefore the need for developing power saving measures in the access network is crucial.

Energy efficiency measures may start by carefully choosing the access network architecture. It is essential to also consider the number of users and transmission speed in the comparison. Other relevant aspects are the day/night traffic fluctuations, bursting traffic nature, use of low-power modes, adapting the line rate and switching off inactive equipment.

In addition to the ICT effort to become greener another measure to reduce the amount of emissions produced, but still produce enough energy to power the network systems is to use renewable energy. The most popular forms of renewable energy generation are by solar and wind energy. As most of the access network density are in the metropolitan area, solar-based solution could be implemented aimed at providing sufficient power for the wired access network as a main source of power rendering the mains to stay as a backup, while having a significant reduction in $CO_2$ emissions [10][11].

## 3. TOWARD CLEAN-SLATE DESIGN OF NGOA

It is clear that the current-generation optical access cannot meet the requirements discussed in Section 2, which forces us to do research into the clean-slate design of NGOA. Here we describe our achievements so far and plans for the future in this study.

### 3.1 A New Comparison Framework

We proposed an extended Equivalent Circuit Rate (ECR) framework for a quantitative comparison of various NGOA architectures in [6], where we compare the user-perceived performance of a candidate architecture to that of a point-to-point reference architecture. In case of shared architecture like TDM-PON, because of contention for a feeder capacity among multiple ONUs and for a distribution capacity among multiple users connected to the same ONU, we can expect that their share of capacity cannot be greater than the minimum of feeder and line rates. Therefore the user-perceived performance would be similar to that of a reference architecture with a line rate equal to or less than the minimum of the feeder and the distribution rates of the candidate architecture.

In [6], we also proposed a systematic ECR calculation procedure which is based on user-perceived performances including average web page delay and decodable frame rate of streaming video as indirect measures of QoE. We adopt non-inferiority testing [12] in comparing the measures to take into account variability in the data from simulation runs.

Currently, we are working toward the extension of the proposed ECR calculation procedure to take into account multiple QoE measures in an integrated way based on Intersection-Union Testing (IUT) and non-parametric version of non-inferiority testing [13].

### 3.2 Implementation of a Virtual Test Bed

Traditionally, research in optical networking has been segmented and quite specialized in certain parts of the whole protocol stack: For instance, most work in the area of optical access has focused on technical issues in physical and data link layers only, and the performance evaluations also have been limited to those two layers and hardly taken into account the upper layers and application- and user-level performances. On the other hand, many of the study on services and applications, network traffic, and high-layer protocols do not consider much the lower layer issues in their investigations. Given the several candidate NGOA architectures & protocols and the recent availability of High-Performance Computing (HPC) clusters & Cloud Computing enabling researchers to carry out a series of large-scale network simulations in a realistic environment, we do believe that now is the time to develop a virtual test bed, together with the comparative analysis framework discussed in Section 3.1, to assess the performances of a network architecture based on the end-user experiences taking into account the impact from the whole network protocol stack and applications as well as a network architecture itself. In this way, we can provide right information on the network architecture back to both end-users and network operators. As part of this virtual test bed for NGOA research initiative, we have implemented detailed simulation models based on OMNeT++ and INET framework [14], which provide models for end-user applications as well as a complete TCP/IP protocol stack [15]. The initial results from the implemented simulation models and the comparison framework have been already reported in [6,16].

As shown in Figure 3, however, a user-level behavioural model is still missing in the current implementation. As for this end-user behaviour, we will first build a demographic and behavioural user profile by focusing on groups for initial exploration and surveying large-scale data collection; because there is no NGOA network deployed now, the use of demographic and behavioural profile obtained from the survey for the architectural study is the only practical option. Then we will build end-user behavioural models governing underlying application-level traffic models based on the developed profile, which can capture the difference between business and residential users and temporal aspects of end-user behaviours.

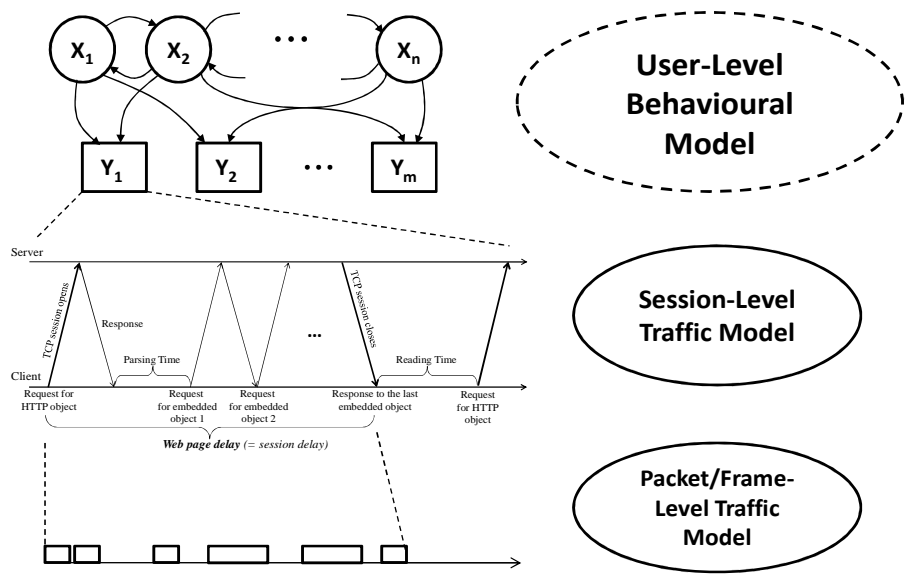

*Figure 3. Multi-level traffic modelling and generation.*

## 4. SUMMARY

We have shown why the clean-slate design of NGOA is unavoidable through discussions on major requirements for NGOA architectures and described the current status of the development of a new comparison framework and the implementation of virtual test bed for the clean-slate design of NGOA.